Direct observation of the compression behavior of polystyrene microbeads in a diamond anvil cell

Haruto Moriguchi, Ken Niwa*, Masashi Hasegawa, Yusuke Koide, Takato Ishida, Takashi Uneyama, and Yuichi Masubuchi*

Department of Materials Physics, and Research Center for Crystalline Materials Engineering, Nagoya University

4648603, Nagoya, Japan

*Corresponding authors: niwa@mp.pse.nagoya-u.ac.jp; mas@mp.pse.nagoya-u.ac.jp

Ver Jul. 8th 2025

**Abstract**

The pressure dependence of the bulk modulus of glassy polystyrene (PS) was measured in the relatively high-pressure regime, up to 6 GPa, at ambient temperature. For the measurements, PS microbeads were immersed in a pressure medium consisting of a mixture of methanol and ethanol, and the sample was placed in a diamond anvil cell capable of generating high and hydrostatic pressure. The volume change of the PS beads was observed under an optical microscope. The results demonstrated that the volume change in this study is consistent with an equation of state determined from the earlier studies in the low-pressure range up to 0.2 GPa. The bulk modulus was obtained as the derivative of the microbead volume with respect to pressure, and compared with the earlier data obtained from Brillouin spectroscopy.

**Introduction**

The change in the specific volume of polymers with temperature and pressure has been extensively measured and discussed since the early times of polymer science[1]. For instance, Adams and Gibson[2] reported the compressibility of a commercial rubber in 1930 as a function of pressure at various temperatures, thereby providing the equation of state. Spencer and Gilmore[3] measured it for polystyrene melts in 1949. Further studies were conducted to measure the specific volume as a function of pressure and temperature for various polymers, with a primary focus on the glass transition and crystallization [4–9].

Nowadays, commodified equipment for dilatometric measurement, referred to as the PVT (pressure-volume-temperature) apparatus, is widely used. The upper limit of the PVT measurement is typically 0.2 GPa, which is related to the processing conditions in the industry.

Although not frequently investigated, measurements for polymers under higher pressure have also been conducted with diamond anvil cells (DAC)[8,10,11]. DAC has been widely used to achieve high-pressure environments, typically up to several GPa. With DAC, Tanaka and Maeda[12] performed dilatometric measurements on small particles of Se glass and InSb crystalline floating in a mixture of methanol and ethanol, combining optical microscopy and image analysis. Tanaka and Shimakawa[13] applied this technique to amorphous phenyl-methyl polysilane films with a molecular weight of 150k. However, no report can be found for other polymers in this direction. Note that Herbst et al.[14] employed a similar experimental setup in which PS microbeads dispersed in methanol were placed in a DAC. They conducted dynamic light scattering measurements to observe the diffusion of the microbeads and discuss the viscosity of the medium as a function of pressure. In the analysis, they considered the pressure dependence of the microbead volume. However, they did not survey the change in the size and shape of the microbeads, and they used data calculated by the equation of state for their analysis.

Apart from dilatometric studies, the elastic properties of some polymers were discussed by combining Brillouin spectroscopy with DAC[15–19]. For instance, Lee et al.[19] reported the bulk, Young's, and shear moduli of some glassy polystyrene samples. Since bulk modulus is the derivative of volume, dilatometric behavior of the material can be discussed. However, the calculation of moduli requires the ratio of the isobaric and the isochoric specific heats, in addition to the sound velocities. Due to technical difficulties, Lee et al.[19] assumed this value to be unity, irrespective of pressure. This assumption has not been validated. Another problem in Brillouin spectroscopy is that the specimen is compressed without a pressure medium that disturbs the phonon. Such direct compression may not achieve hydrostatic conditions. Indeed, Lee et al.[19] mentioned that a pressure gradient existed in their DAC. We also note that due to the viscoelasticity of polymers, the mechanical response depends on the frequency of the probe. Indeed, an earlier study[20] on the measurement of Young's modulus of some polymers under pressure using ultrasound reported that the obtained values are larger than those from elongational measurements due to the high sound frequency employed.

In this study, following the dilatometric experiments with DAC[12,13], we measured the volume of spherical polystyrene (PS) microbeads under high pressure, reaching up to 6 GPa. We employed a DAC apparatus to achieve high pressures and placed PS beads immersed in a liquid medium in the chamber to observe their diameter as the pressure varied. We converted the obtained data to the bulk modulus and compared it with the value measured by Brillouin measurements [19]. The results revealed that the bulk modulus of PS in the two measurements differed from each other. We discussed possible reasons for the discrepancy. Details are shown below.

**Materials and Methods**

A modified Mao–Bell type DAC[22] with a culet diameter of 450 μm was used in this study. The sample chamber with a central hole of 250 μm in diameter was prepared in a stainless-steel gasket of initial thickness 250 μm.

PS microspheres with a diameter of 45 μm were purchased from Polysciences, Inc., and used as supplied. At each run, several PS beads (fewer than 5) were placed in a DAC chamber. The chamber was filled with a 4:1 weight ratio mixture of methanol and ethanol. Both liquids were purchased from Wako and used as supplied. The volume ratio is nearly identical to the weight ratio. As the pressure probe, a few ruby chips, approximately 100 μm in size, were also dispersed in the sample liquid. [21,22] Filling the sample and pressure medium into the sample chamber of the DAC was performed under a stereomicroscope (SMZ-U, Nikon).

Through the diamonds, optical microscope observation was carried out under a microscope (MM-400, Nikon) with a 50x magnification of the objective lens. The optical images were then acquired by a CCD camera (DS-Fi1, Nikon) for further analysis, as shown later. A DPSS laser with a wavelength of 532 nm was irradiated onto the ruby tip to induce fluorescence[23], the fluorescence peak in the 690–700 nm range was determined by Gaussian fitting and subsequently converted to pressure using the Piermarini equation[24]. We have confirmed that the half-width and bimodal shape of the ruby fluorescence were well preserved in the measurements of this study. The thickness of the gasket after experiment was confirmed to be at least 50 μm, which was larger than the diameter of the PS bead.

Figure 1 (a) shows a typical snapshot of a microbead under pressure at 6 GPa. All the observed beads maintained a spherical shape in the examined pressure range, and neither breakage, melting, dissolution, nor anisotropic deformation by direct compression with the anvils was detected, at least in a visible manner. The images were taken and analyzed for steady states, which were achieved after a relaxation time of less than 2 minutes when pressure was varied.

From such images, the radius of the microbead was extracted by fitting the observed bead shape to a circle after image processing, containing masking of the transmitted light through the bead (Fig 1 (b)), binarization (Fig 1 (c)), and edge detection (Fig 1 (d)). The curve thus obtained was fitted as an arc, and its radius was regarded as that of the bead. Assuming a spherical shape, the volume of the PS bead was then estimated based on the extracted radius. OpenCV[25] and ImageJ[26] were employed, and Otsu's thresholding method[27] and the Canny algorithm[28] were used for binarization and edge detection, respectively.

The pressure was varied from ambient conditions to 6 GPa at a rate of 0.5 GPa/min. Under these conditions, the volume of each bead reached a practically steady value at every realized pressure within experimental uncertainty, even though the longest relaxation time of PS glass is expected to be much longer. The pressure-dependent volume changes were examined for five individual PS beads. As shown in Fig. 1(a), some PS beads appear to be in shallow contact with one another. However, since each bead was visually confirmed to shrink isotropically and independently, we neglect the effect of such contact between beads.

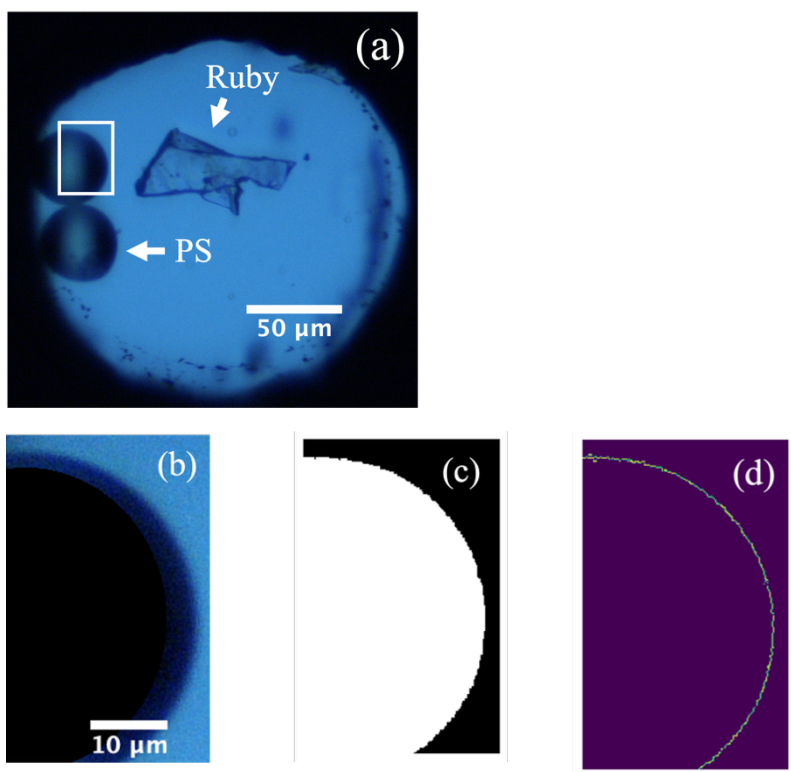

**Figure 1** (a) A Typical image of the bead in the DAC under pressure at 6 GPa, and the processed images for the detection of the bead radius, containing (b) masking of transmitted light through the bead, (c) binarization, and (d) edge detection.

**Results and Discussion**

Figure 2 shows an example of the volume response of a PS bead during a pressurized–depressurized cycle up to 6 GPa. As reported in conventional PVT studies, PS volume decreases with increasing pressure. However, the attained compression is significantly higher due to the high pressure in the employed DAC; the volume is reduced to 70% of its initial state, whereas compression in conventional PVT measurements typically reaches up to 95% under 0.2GPa[7]. No hysteresis was observed under these conditions, indicating the reversibility of the volume change of PS microspheres.

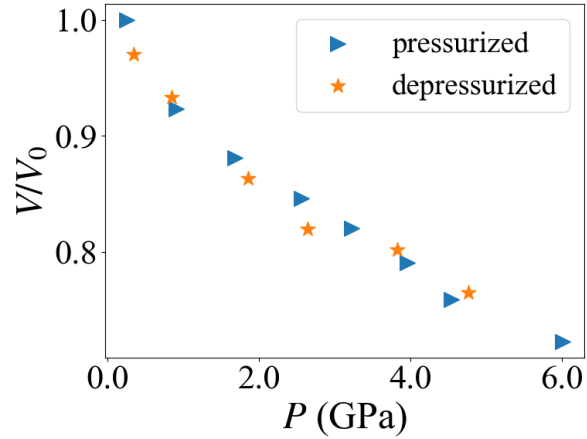

**Figure 2** Volume change of a PS bead under pressurization (triangle) and depressurization (star). The pressure change rate was 0.5 GPa/min. The volume $V$ was normalized by the value $V_0$ at ambient pressure (~0.1GPa).

Figure 3 illustrates the variation among different PS beads under pressurization. The rate of change of pressure was slower than 0.5 GPa/min for all cases. Overall, a consistent monotonic volume shrinkage was observed for all the examined cases. Differences among the runs and samples exhibit inevitable errors in image analysis and technical difficulties with pressure control. Nevertheless, we observe no apparent discontinuity or discontinuous change of slope. This behavior implies that the PS beads exhibit neither a glass transition nor a phase transition, at least within the examined range.

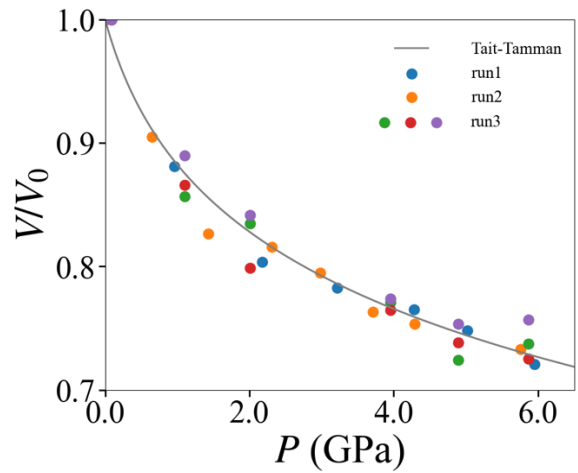

**Figure 3** Volume change of several different PS beads under pressurization. Symbols in different colors exhibit different runs. The solid curve represents Eq. 1 with $B = 0.49$ and $C = 0.106$. Each labeled group (Run 1, Run 2, Run 3) denotes an independent pressurization experiment. Multiple plots in different colors within a run indicate different PS beads measured in the same experiment.

In Fig. 3, the obtained data are compared with the Tait-Tamman equation[29]:

$$\frac{V}{V_0} = 1 - C \log\left(1 + \frac{P}{B}\right). \quad (1)$$

The solid curve shows the fitting to our data with the parameters B = 0.49 ± 0.18 GPa and $C$ = 0.106 ± 0.016, for which the errors correspond to the scattering of data among different beads and experimental runs. The determined parameters are in reasonable agreement with the earlier results obtained from the low-pressure PVT data[7], B = 0.312 GPa and C = 0.0894, implying consistency between our measurement and the earlier PVT experiment.

Owing to the fitting of the data in Fig. 3 to eq. 1, let us discuss the bulk modulus $K = -V(dP/dV)$. Figure 4 shows $K(P)$ curve thus obtained in comparison to the Brillouin spectroscopy data reported by Lee et al. earlier[19] for PS with a molecular weight $M_w$ of 350,000. Here, we exhibit the uncertainty of our fitting by the gray area. The Brillouin spectroscopy data are lower than our curve, falling outside our uncertainty range.

A possible reason for this discrepancy in $K(P)$ is the difference in the measured systems and the measurement techniques. For the system, we employed the 4:1 methanol-ethanol mixture as the medium, whereas Lee et al. [19] directly pressurized their PS samples. Concerning the use of the 4:1 methanol-ethanol mixture, Casado et al.[30] reported the PV behavior at ambient temperature using a DAC. The specific volume of the liquid reduces to 60% at 5 GPa from the value under ambient pressure. Although this compression is more significant than that of PS, the mixture has sufficient hydrostatic pressure [30] within the examined pressure range. Swelling of the PS beads due to pressure-induced mixing between PS and the liquid may occur; however, the authors found no literature on this phenomenon. In contrast, the direct compression is free from the effect of the pressurized medium. However, Lee et al. [19] reported that the direct compression involved a pressure gradient and did not achieve the hydrostatic condition. The other difference is the molecular weight of PS. The data by Lee et al. [19] in Fig. 4 are for PS with a molecular weight $M_w$ of 350,000, and they reported that the bulk modulus depends on the molecular weight. Although the molecular weight of our PS beads is unknown, the variation of $K$ for a range of $M_w$ from 3,700 to 350,000 is much less than the discrepancy in Fig. 4.

The other issue is the ratio of the isobaric and the isochoric specific heats. As mentioned in the introduction, this value is required for converting sound velocities to moduli in Brillouin spectroscopy. Although Lee et al.[19] assumed that this value is unity independent of pressure, their cited literature[31] tells that the value ranges up to 1.2. We note that this difference is not negligible; indeed, if 1.2 is employed, the calculated modulus by Lee et al. [19] increases and falls within our experimental uncertainty.

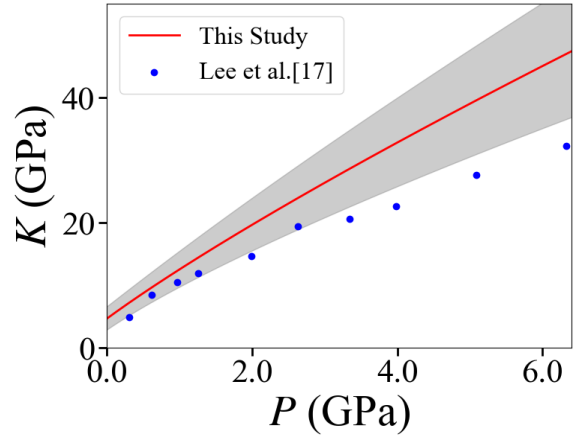

**Figure 4** Pressure dependence of bulk modulus ($K$) obtained from Eq. 1 with = 0.492 GPa and $C = 0.106$, and the literature data from Brillouin spectroscopy [19] at $M_w$ =350,000 (symbol). The hatched area displays the uncertainty of our experimental data.

For completeness, let us discuss the viscoelastic nature of the material. As established, the mechanical properties of polymers, such as modulus, depend on the frequency of applied mechanical stimulation, reflecting molecular motion on different length scales; Young's modulus and shear modulus of glassy polymers increase with increasing frequency[32]. This frequency dependence manifests as differences in modulus reported by various experimental methods. For instance, Masubuchi et al. [20] measured Young's modulus of a few glassy polymers using the velocity of transmitted ultrasound. The Young's modulus of polymethylmethacrylate is 6 GPa at an ultrasound frequency of 5 MHz. This modulus is higher than the value of 3 GPa obtained from stretching mechanical measurements at a frequency of $10^{-3}$ Hz, reflecting the difference in the frequency of the applied mechanical deformations. Although not frequently discussed, the bulk modulus is also viscoelastic [33,34], and a similar frequency effect is likely to exist. However, our results exhibit the opposite trend; the bulk modulus from our measurements is higher than that from Brillouin spectroscopy, even though the analyzed phonon frequency (around 1 GHz) is significantly higher than the characteristic compression rate of our measurements.

**Conclusions**

To investigate the pressurized behavior of glassy PS, we conducted a direct observation of PS microbeads dispersed in a 4:1 methanol-ethanol mixture as a pressure-transmitting medium in a DAC. Under a pressure of up to 6 GPa, with a pressurization rate of 0.5

GPa/min, the compression of microbeads was isotropic and reversible, without breakage or melting, at least apparently. The observed pressure dependence of the bead volume is consistent with the Tait-Tamman equation of state for the parameter set determined by the conventional PVT measurements in the literature. The bulk modulus was then calculated from the PV relation and compared with that obtained from Brillouin spectroscopy reported earlier. The comparison revealed that the bulk modulus in our measurement is larger than that obtained from Brillouin spectroscopy. The reason for this discrepancy is unknown, although we have discussed possible reasons related to the difference in measurement techniques. Further studies, varying temperatures, and conducting structural analysis may offer additional insight.